\documentclass[twocolumn,conference,a4paper,11pt]{IEEEtran}
\usepackage{amssymb,amsmath,epsfig,latexsym,graphicx,bm}
\usepackage{epstopdf}
\DeclareGraphicsExtensions{.eps}
\usepackage{booktabs}
\usepackage{multirow}
\usepackage{float}
\usepackage{tabularx}
\usepackage{cite}
\usepackage{flushend}

\IEEEoverridecommandlockouts

\begin{document}
%

%

\title{A Variational Step for Reduction of Mixed Gaussian-Impulse Noise from Images}

\author{\IEEEauthorblockN{Mohammad~Tariqul~Islam,$^{1}$ Dipayan Saha,$^{1}$ \\S.~M.~Mahbubur~Rahman,$^{1}$ M.~Omair~Ahmad,$^{2}$ and M.~N.~S.~Swamy$^{2}$}
\IEEEauthorblockA{$^1$Department of Electrical and Electronic Engineering\\
Bangladesh University of Engineering and Technology, Dhaka-1205, Bangladesh}
\IEEEauthorblockA{$^2$Department of Electrical and Computer Engineering\\
Concordia University, Montr\'{e}al, QC, H3G 1M8, Canada\\
Email: tariqul@eee.buet.ac.bd, dipayansahabd@gmail.com,\\ mahbubur@eee.buet.ac.bd, omair@ece.concordia.ca, swamy@ece.concordia.ca}}



\IEEEpubid{\begin{minipage}{\textwidth}\ \\[12pt]
  Accepted in ICECE, Dhaka, Bangladesh, 2018
\end{minipage}}

%
%

\maketitle

\begin{abstract}

Reduction of mixed noise is an ill posed problem for the occurrence of contrasting distributions of noise in the image. The mixed noise that is usually encountered is the simultaneous presence of additive white Gaussian noise (AWGN) and impulse noise (IN). A standard approach to denoise an image with such corruption is to apply a rank order filter (ROF) followed by an efficient linear filter to remove the residual noise. However, ROF cannot completely remove the heavy tail of the noise distribution originating from the IN and thus the denoising performance can be suboptimal. In this paper, we present a variational step to remove the heavy tail of the noise distribution. Through experiments, it is shown that this approach can significantly improve the denoising performance of mixed AWGN-IN using well-established methods.

\end{abstract}


%

\section{Introduction}

Image denoising is a fundamental problem in image processing. The 
physical properties of imaging devices as well as faulty sensors 
and transmission equipments are primarily responsible for noise 
in images~\cite{rabie2004adaptive}. Two types of noise which are 
often encountered in practice are the additive white Gaussian 
noise (AWGN) and the impulse noise (IN). AWGN is commonly 
introduced by the temperature of the sensor and the level of illumination  
in the environment  that corrupts every pixels, whereas IN is caused 
by faulty sensor triggers or transmissions corrupting a certain 
image pixels~\cite{gonzalez2008digital}. Due to the common 
origin, these noises often occur simultaneously in practice 
which includes digital photography~\cite{grou2009random}, 
tomography~\cite{norose2015impulsect} and thermal 
imaging~\cite{qi2011wavelet}. This mixed noise removal has 
been studied fairly well in the past~\cite{peng1994fuzzy, cai2008two, huang2009fast, cai2010fast, xiong2012universal, jiang2014mixed, huang2017mixed, islam2018mixed}.

Removing mixed AWGN-IN is relatively difficult due to the unique nature 
of the AWGN and IN. In general, the order statistics filters are 
effective in reducing impulse noise whereas the additive filters are 
successful in reducing the AWGN~\cite{peng1994fuzzy}. Thus the common 
approach to tackle the mixed AWGN-IN is to employ an order statistics 
filter which detects and removes the IN and then use a smoothing 
algorithm to remove the remaining residual noise. It is natural to consider the 
residual noise as Gaussian like and use a AWGN denoiser to remove the 
residual noise. A common benchmark in the literature uses a rank order 
filter (ROF) such as adaptive median filter (AMF)~\cite{hwang1995adaptive} 
for salt and pepper impulse noise, followed by a Gaussian denoiser such 
as block matching and 3D filtering (BM3D) AWGN denoiser~\cite{dabov2007image}. 
In a similar fashion, Xiong {\emph{et al.}}~\cite{xiong2012universal} 
recommended a robust outlyingness ratio (ROR) statistics of neighboring 
pixels to detect IN, and then employed non local means to remove mixed 
noise from images. The next class of methods use rank order filter 
to detect and remove impulse noise and then employ an optimization 
technique to obtain noise free images. The optimization techniques 
explored under this framework include $l_1$ norm~\cite{cai2008two,cai2010fast} 
and total variation norm~\cite{huang2009fast}. Non-local 
regularization, in the second phase, has generated considerable 
success. In order to unify the framework of impulse detection 
and noise reduction, Jiang~\emph{et al.}~\cite{jiang2014mixed} 
employed weighted encoding of the mixed noise to denoise the images. 
The method requires initialization of the estimated noise free image, 
which is performed employing a ROF on the noisy image. In a similar 
framework, Huang \emph{et al.}~\cite{huang2017mixed} employed 
Laplacian scale mixtures modeling to fit the IN and employed 
non-local low rank regularizer to reduce mixed AWGN-IN. 
Recently, variational impulse removal followed by CNN has 
been employed in~\cite{islam2018mixed}.


\begin{figure*}[!t]
\centering
\includegraphics[width=8cm]{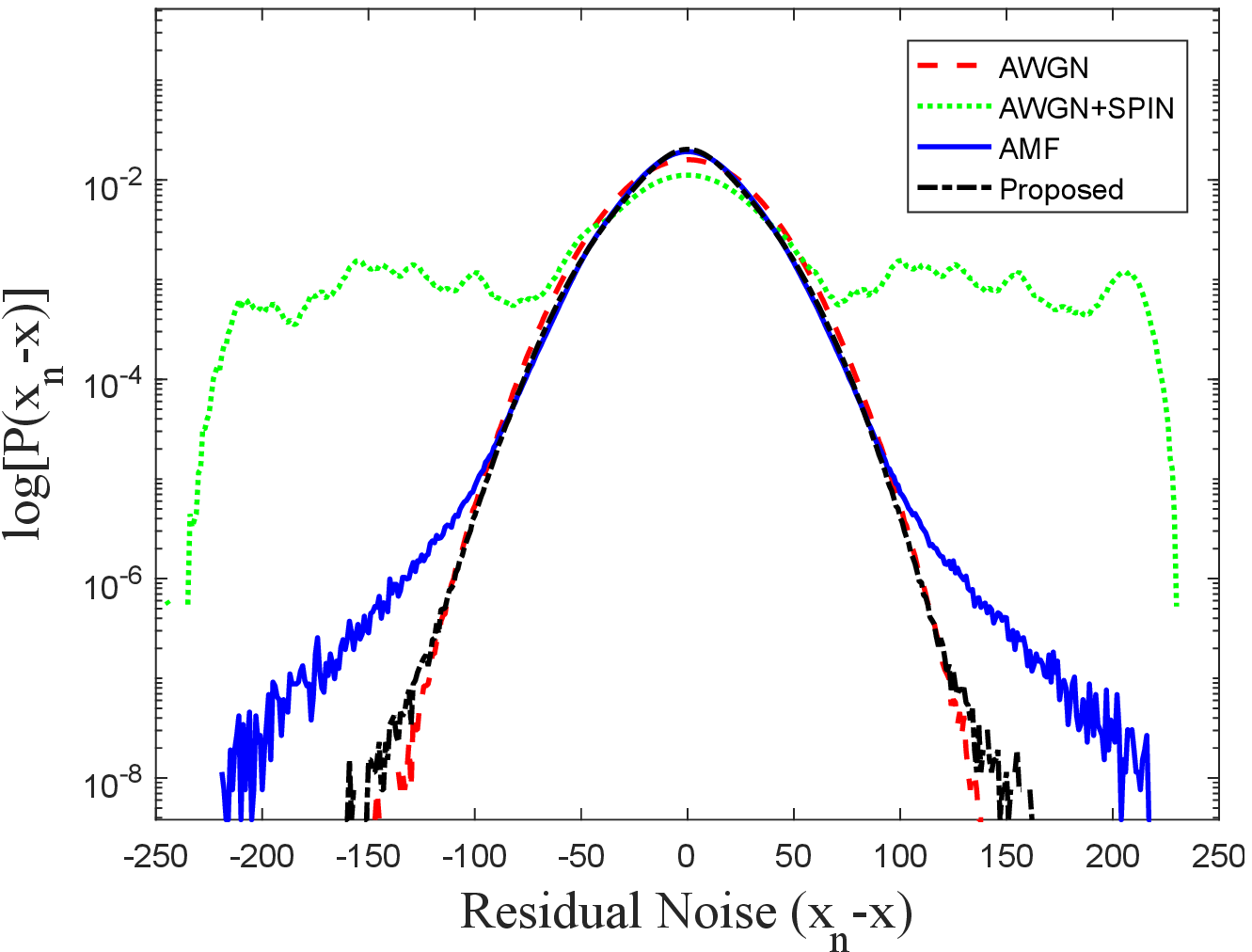}
\hspace{1cm}
\includegraphics[width=8cm]{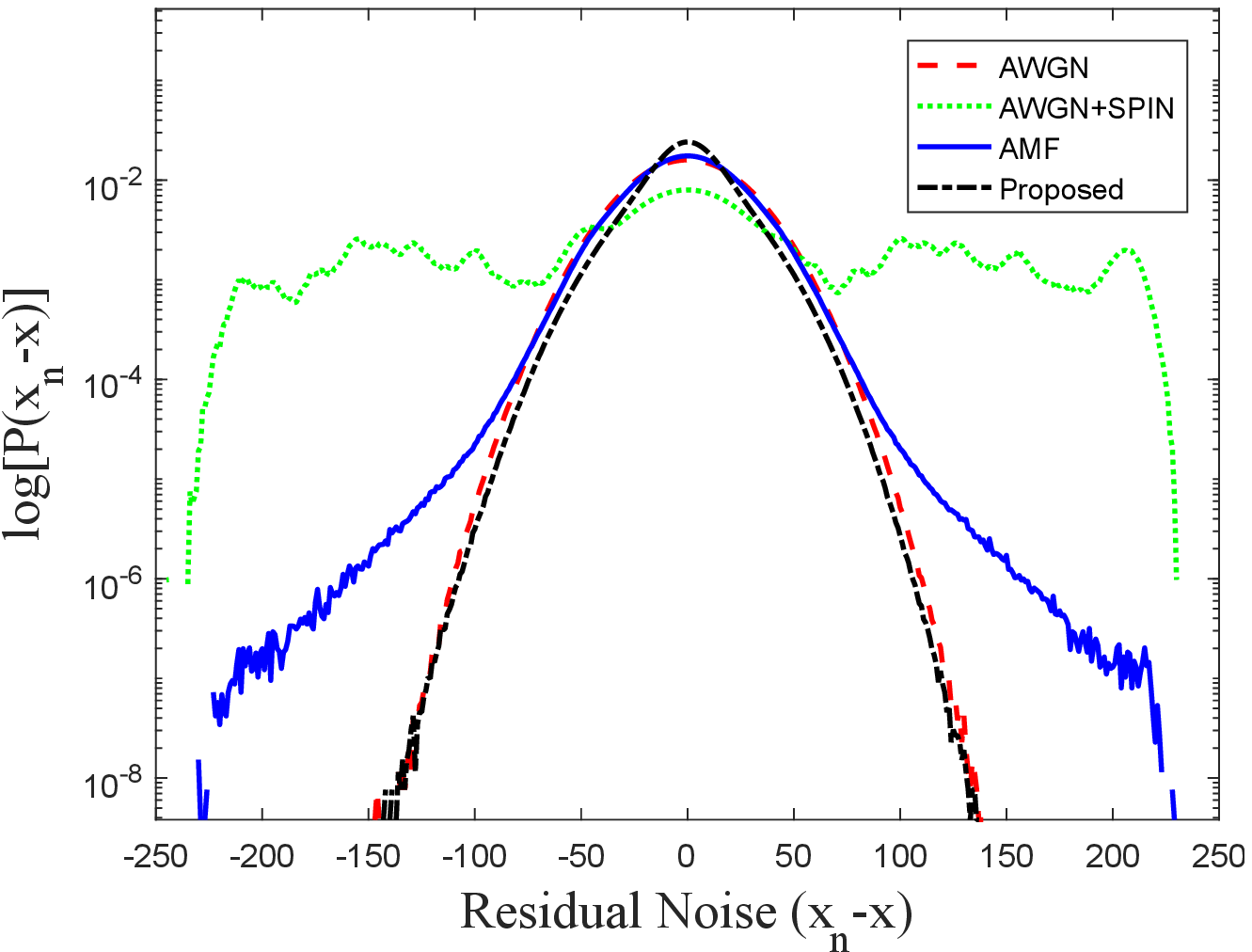}\\
(a)\hspace{8cm}(b)\\
\caption{\label{fig:residual_distribution}Empirical distribution in log scale considering residual noise simulated on \textit{Lena} image when image is corrupted by AWGN and mixed AWGN+SPIN, and then IN is removed using AMF and the proposed method. The parameters of two scenarios of AWGN+SPIN are (a) $\sigma=25$, $p=30\%$, (b) $\sigma=25$, $p=50\%$.}
\end{figure*}

From the discussion, it is evident that the methods generally require impulse detection and removal step before it can be tackled for overall noise removal. This impulse removal step, generally, can be considered as a ``Gaussianization" step, which removes the heavy tail caused by the IN. Since, the AWGN removal process is pretty well established, if efficient algorithms to remove the heavy tail are available, it will be possible to increase the overall denoising performance. With this motivation, in this paper, a variational approach with local regularization is proposed to reduce the heavy tail of the noise. We show that performance of the existing methods can be significantly improved using our proposed approach by experimenting on several well established methods. As a result, this method can be employed as an important step in mixed AWGN-IN removal algorithm.




\section{Problem Formulation}

Under the mixed AWGN-IN degradation model, the observation $x_n \in R^{M \times N}$ of 
the noisy image can be modeled as a function of its noise free 
version $x \in R^{M \times N}$ as~\cite{jiang2014mixed}
\begin{align}
x_n = f(x)
\end{align}
where $f(\cdot)$ is the degradation function. In this paper two types of degradation are considered: 1) mixed AWGN and salt and pepper impulse noise (SPIN) and 2) mixed AWGN, SPIN and random valued impulse noise (RVIN). Let a pixel of the noisy image be denoted as $x_n(i,j)$. For the case when an image is corrupted by AWGN alone, the noisy pixel is given by,
\begin{align}
x_n(i,j)=x(i,j)+\nu(i,j)
\end{align}
where $\nu(i,j)$ is a sample of i.i.d. zero-mean Gaussian
distribution with standard deviation $\sigma$. 
Let the dynamic rage of the image pixels be within the range $[d_{max},d_{min}]$.
In such case, the SPIN originates when an image pixel is stuck either 
in the maximum pixel value $d_{max}$ with probability $p/2$ or the minimum pixel value $d_{min}$ with probability $p/2$, where $p\leq 1$.
In a mixed AWGN+SPIN scenario, the pixel is contaminated by AWGN with 
a probability $(1-p)$.
Similarly, an image pixel $x_n(i,j)$ is corrupted with RVIN when it 
gets stuck to a random value $d(i,j)$ with
probability $r$ $(r\leq 1)$. The value $d(i,j)$ is uniformly
distributed in the dynamic range
$\left[d_{min},d_{max}\right]$.
Using this definition of SPIN and RVIN, the general case 
of an image pixel $x_n(i,j)$ corrupted by mixed AWGN-IN
can be given by~\cite{jiang2014mixed}
\begin{align}
&x_n(i,j)=\nonumber\\
&\left\{\begin{array}{cl}
d_{min}&\textrm{with probability}~p/2\\
d_{max}&\textrm{with probability}~p/2\hspace{0.5cm}.\\
d(i,j)&\textrm{with probability}~r(1-p)\\
x(i,j)+\nu(i,j)&\textrm{with probability}~(1-p)(1-r)
\end{array}\right.\label{eq_rvin}
\end{align}
If $r=0$, the noise is mixed AWGN+SPIN, otherwise the noise is mixed AWGN+SPIN+RVIN.
The purpose of denoising is to estimate the noise free image $x$ from the corresponding noisy observation $x_n$.

\section{Proposed Method}

In the proposed method, we focus on removing impulse noise to remove the heavy tail caused by it. In this light, we took inspiration from the impulse removal and detection approach provided in~\cite{cai2010fast} and employed an $l_1$-norm based regularization approach to lessen the effect of impulse noise.

Let $z\in\mathbb{R}^{M\times N}$ be the image obtained by the rank order filtering operation on the mixed noise corrupted image $x_n$. The filtered image $z$ is employed to determine the noise candidates by
\begin{align}
\mathcal{N} = \{(i,j)\in\mathcal{A} | z(i,j)\neq x_n(i,j)\}
\end{align}
where $\mathcal{N}$ is the set of pixels locations corrupted by impulse noise, $\mathcal{A}$ is the set of all observed pixels locations. A proper choice of impulse detector should detect most of the noisy pixels successfully. For SPIN, the set $\mathcal{N}$ can be the set of locations of $d_{max}$ and $d_{min}$. Thus, locations of the pixels that are uncorrupted by impulse noise are defined as
\begin{align}
&\mathcal{U} = \mathcal{A}\setminus\mathcal{N}
\end{align}

The optimization is performed only on these uncorrupted image pixels. The resultant ill-posed inverse problem is solved by using a variational method and requires minimization of the convex function given by~\cite{cai2010fast}
\begin{align}
&\sum_{(i,j)\in\mathcal{A}} \mathcal{\chi}(i,j) |x(i,j)-x_n(i,j)| \nonumber \\
&+ \beta \sum_{(i,j)\in\mathcal{A}} \sum_{(k,l)\in\mathcal{V}_{i,j}}
| x(i,j)-x(k,l) |
\end{align}
where the first term in the function is the $l_1$ norm, the second term is an edge preserving local regularizer, $\chi$ is the characteristic function of the set $\mathcal{U}$, $\beta$ is the regularizing parameter and $\mathcal{V}_{i,j}$ is the set of four neighboring pixels of the pixel at $(i,j)$ for local regularization. The function $\chi$ is given by
\begin{align}
&\chi(i,j)=
\left\{\begin{array}{cl}
1 &\textrm{if } (i,j)\in \mathcal{U}\\
0 &\textrm{otherwise}
\end{array}\right.
\end{align}

Following~\cite{vogel1996iterative}, the function can be optimized using fixed point iteration by introducing a weak smooth regularizer given by
\begin{align}
\mathcal{L}(x) = &\sum_{(i,j)\in\mathcal{A}} \sqrt{\mathcal{\chi}(i,j) [x(i,j)-x_n(i,j)]^2 + \eta} \nonumber \\
& + \beta \sum_{(i,j)\in\mathcal{A}} \sum_{(k,l)\in\mathcal{V}_{i,j}}
\sqrt{[ x(i,j)-x(k,l) ]^2 + \eta}
\end{align}
which can be differentiated and equated to zero to obtain a solution. Given the solution of $x$ at $(p-1)$th iteration, the solution at $p$th iteration can be computed by solving the following linear equation~\cite{cai2010fast}
\begin{align}
\frac{\chi \circ [x^{p}-x_n]}{\sqrt{[x^{p-1}-x_n]^2+\eta}} + \beta G^{*} \frac{[Gx^{p}]}{\sqrt{[Gx^{p-1}]^2+\eta}}=0
\end{align}
where $\circ$, $[\cdot]^2$, and $\frac{\cdot}{\cdot}$ are elementwise multiplication, square and division, respectively, $G$ is the difference matrix such that $Gx(ij,kl)=x(i,j)-x(k,l)$ for $(i,j)\in\mathcal{A}$ and $(k,l)\in\mathcal{V}_{i,j}$, and $G^{*}$ is the adjoint matrix of $G$. A good choice of $\beta$ makes it an efficient process to remove the heavy tail.

Fig.~\ref{fig:residual_distribution} shows the residual noise of the \textit{Lena} image in four scenarios: the image is corrupted by AWGN only, it is corrupted by mixed AWGN-IN, the IN removal process using AMF and the proposed variational step, for different noise parameters. The proposed method is employed by $\beta=0.0002$ and a tolerance of $0.001$ in the numerical solution. It can be observed from the figure that the process effectively reduces the heavy tail of the residual noise. The effect can be clearly observed from the logarithmic scale of the distributions. 

A typical mixed AWGN-IN removal algorithm has two primary step. First, the IN is removed using an ROF and then, the residual noise is removed using another algorithm. Under the proposed method, the ROF is followed by this variational step and then the denoiser follows.

\begin{table*}[t]
  \centering
  \caption{\label{tab:spin}Comparison of ROF+BM3D, WESNR, and LSM-NLR methods between default settings and proposed modification for mixed AWGN+SPIN reduction in terms of PSNR and SSIM.}%
  \scalebox{0.75}{
    \begin{tabular}{cccccccccccc}
    \toprule
    \multicolumn{1}{c}{Image Name} & \multicolumn{1}{c}{Noise Parameter} & \multicolumn{1}{c}{Metric} & \multicolumn{1}{c}{ROF+BM3D~\cite{dabov2007image}} & \multicolumn{1}{c}{MBM3D} & \multicolumn{1}{c}{\% Increase} & \multicolumn{1}{c}{WESNR~\cite{jiang2014mixed}} & \multicolumn{1}{c}{MWESNR} & \multicolumn{1}{c}{\% Increase} & \multicolumn{1}{c}{LSM-NLR~\cite{huang2017mixed}} & \multicolumn{1}{c}{MLSM-NLR} & \multicolumn{1}{c}{\% Increase} \\
    \midrule
    \multicolumn{1}{c}{\multirow{4}{*}{Lena}} & \multicolumn{1}{c}{\multirow{1}{*}{$\sigma=25$}} & PSNR & $29.40$ & $30.67$ & $4.32$  & $30.04$ & $30.49$ & $1.51$ & $30.94$ & $31.02$ & $0.25$  \\
\cmidrule{3-12}          & \multicolumn{1}{c}{\multirow{1}{*}{$p=30\%$}} & SSIM & $0.7860$ & $0.8257$ & $5.05$  & $0.8216$ & $0.8272$ & $0.68$ & $0.8401$ & $0.8412$ & $0.13$ \\
\cmidrule{2-12}          & \multicolumn{1}{c}{\multirow{1}{*}{$\sigma=25$}} & PSNR & $27.33$ & $29.38$ & $7.49$  & $28.92$ & $29.64$ & $2.48$ & $29.99$ & $30.37$ & $1.27$ \\
\cmidrule{3-12}          & \multicolumn{1}{c}{\multirow{1}{*}{$p=50\%$}} & SSIM & $0.7040$ & $0.7908$ & $12.33$ & $0.8032$ & $0.8137$ & $1.31$ & $0.8305$ & $0.8326$ & $0.25$ \\
    \midrule
    \multicolumn{1}{c}{\multirow{4}[4]{*}{House}} &  \multicolumn{1}{c}{\multirow{1}{*}{$\sigma=25$}}  & PSNR & $29.72$ & $31.23$ & $5.11$  & $30.48$ & $31.00$ & $1.70$ & $31.56$ & $31.63$ & $0.25$ \\
\cmidrule{3-12}          & \multicolumn{1}{c}{\multirow{1}{*}{$p=30\%$}} & SSIM & $0.7828$ & $0.8239$ & $5.25$  & $0.8296$ & $0.8352$ & $0.68$ & $0.8401$ & $0.8410$ & $0.11$   \\
\cmidrule{2-12}          &  \multicolumn{1}{c}{\multirow{1}{*}{$\sigma=25$}} & PSNR & $27.19$ & $29.69$ & $9.17$  & $29.50$ & $30.24$ & $2.51$ & $30.51$ & $31.18$ & $2.20$ \\
\cmidrule{3-12}          & \multicolumn{1}{c}{\multirow{1}{*}{$p=50\%$}} & SSIM & $0.6859$ & $0.7933$ & $15.66$ & $0.8117$ & $0.8258$ & $1.74$ & $0.8303$ & $0.8379$ & $0.92$   \\
    \midrule

    \multicolumn{1}{c}{\multirow{4}[4]{*}{Boat}} &  \multicolumn{1}{c}{\multirow{1}{*}{$\sigma=25$}}  & PSNR & $27.24$ & $28.44$ & $4.41$  & $27.35$ & $28.02$ & $2.47$ & $28.58$ & $28.74$ & $0.57$ \\
\cmidrule{3-12}          & \multicolumn{1}{c}{\multirow{1}{*}{$p=30\%$}} & SSIM & $0.7205$ & $0.7588$ & $5.32$  & $0.7225$ & $0.7362$ & $1.90$ & $0.7665$ & $0.7682$ & $0.22$\\
\cmidrule{2-12}          &  \multicolumn{1}{c}{\multirow{1}{*}{$\sigma=25$}}  & PSNR & $25.49$ & $27.12$ & $6.39$  & $25.97$ & $27.03$ & $4.10$ & $27.34$ & $27.74$ & $1.45$ \\
\cmidrule{3-12}          & \multicolumn{1}{c}{\multirow{1}{*}{$p=50\%$}} & SSIM & $0.6525$ & $0.7143$ & $9.47$  & $0.6968$ & $0.7056$ & $1.26$ & $0.7326$ & $0.7344$ & $0.25$\\
    \bottomrule
    \end{tabular}%
    }
  
\end{table*}%

\begin{table*}[!htbp]
  \centering
  \caption{\label{tab:rvin}Comparison of ROF+BM3D, WESNR, and LSM-NLR methods between default settings and proposed modification for mixed AWGN+SPIN+RVIN reduction in terms of PSNR and SSIM.}%
  \scalebox{0.75}{
    \begin{tabular}{cccccccccccc}
    \toprule
    \multicolumn{1}{c}{Image Name} & \multicolumn{1}{c}{Noise Parameter} & \multicolumn{1}{c}{Metric}  & \multicolumn{1}{c}{ROF+BM3D~\cite{dabov2007image}} & \multicolumn{1}{c}{MBM3D} & \multicolumn{1}{c}{\% Increase} & \multicolumn{1}{c}{WESNR~\cite{jiang2014mixed}} & \multicolumn{1}{c}{MWESNR} & \multicolumn{1}{p{5.145em}}{\% Increase} & \multicolumn{1}{c}{LSM-NLR~\cite{huang2017mixed}} & \multicolumn{1}{c}{MLSM-NLR} & \multicolumn{1}{c}{\% Increase} \\
    \midrule
    
    \multicolumn{1}{c}{\multirow{2}[2]{*}{Lena}} & \multicolumn{1}{c}{\multirow{1}{*}{$\sigma=10$, $p=25\%$,}} & PSNR  & $32.13$ & $33.47$ & $4.16$ & $32.91$ & $33.01$ & $0.58$ & $33.30$ & $33.59$ & $0.85$   \\
\cmidrule{3-12}          & \multicolumn{1}{c}{\multirow{1}{*}{$r=5\%$}}        & SSIM & $0.8600$ & $0.8734$ & $1.56$ & $0.8705$ & $0.8733$ & $0.32$ & $0.8926$ & $0.8938$ & $0.13$ \\
    \midrule
    
    \multicolumn{1}{c}{\multirow{2}[2]{*}{House}} & \multicolumn{1}{c}{\multirow{1}{*}{$\sigma=10$, $p=25\%$,}} & PSNR & $32.16$ & $33.42$ & $3.91$ & $33.25$ & $33.49$ & $0.69$ & $33.51$ & $33.92$ & $1.21$    \\
\cmidrule{3-12}          &  \multicolumn{1}{c}{\multirow{1}{*}{$r=5\%$}} & SSIM & $0.8596$ & $0.8718$ & $1.42$ & $0.8623$ & $0.8650$ & $0.31$  & $0.8872$ & $0.8889$ & $0.19$   \\
    \midrule

    \multicolumn{1}{c}{\multirow{2}[2]{*}{Boat}} & \multicolumn{1}{c}{\multirow{1}{*}{$\sigma=10$, $p=25\%$,}} & PSNR & $29.14$ & $29.79$ & $2.20$ & $29.64$ & $29.77$ & $0.42$ & $29.99$ & $30.21$ & $0.74$  \\
\cmidrule{3-12}          & \multicolumn{1}{c}{\multirow{1}{*}{$r=5\%$}}  & SSIM & $0.8169$ & $0.8284$ & $1.41$ & $0.8171$ & $0.8207$ & $0.44$ & $0.8407$ & $0.8429$ & $0.26$ \\
    \bottomrule
    \end{tabular}%
    }
  
\end{table*}%

\section{Experiments and Results}~\label{sec:experiments}

\begin{figure*}[!t]
\centering
\includegraphics[height=4cm,width=4cm]{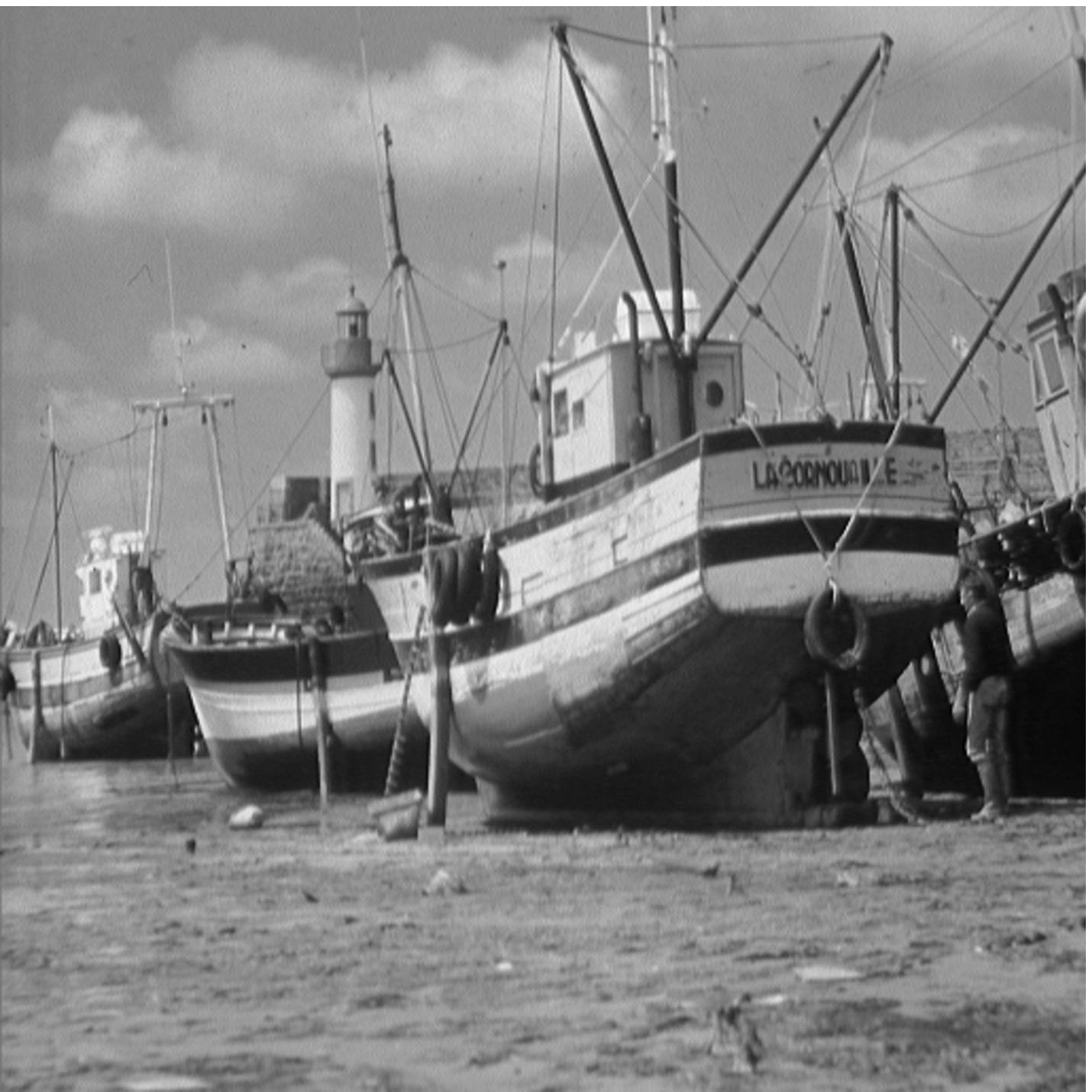}
\includegraphics[height=4cm,width=4cm]{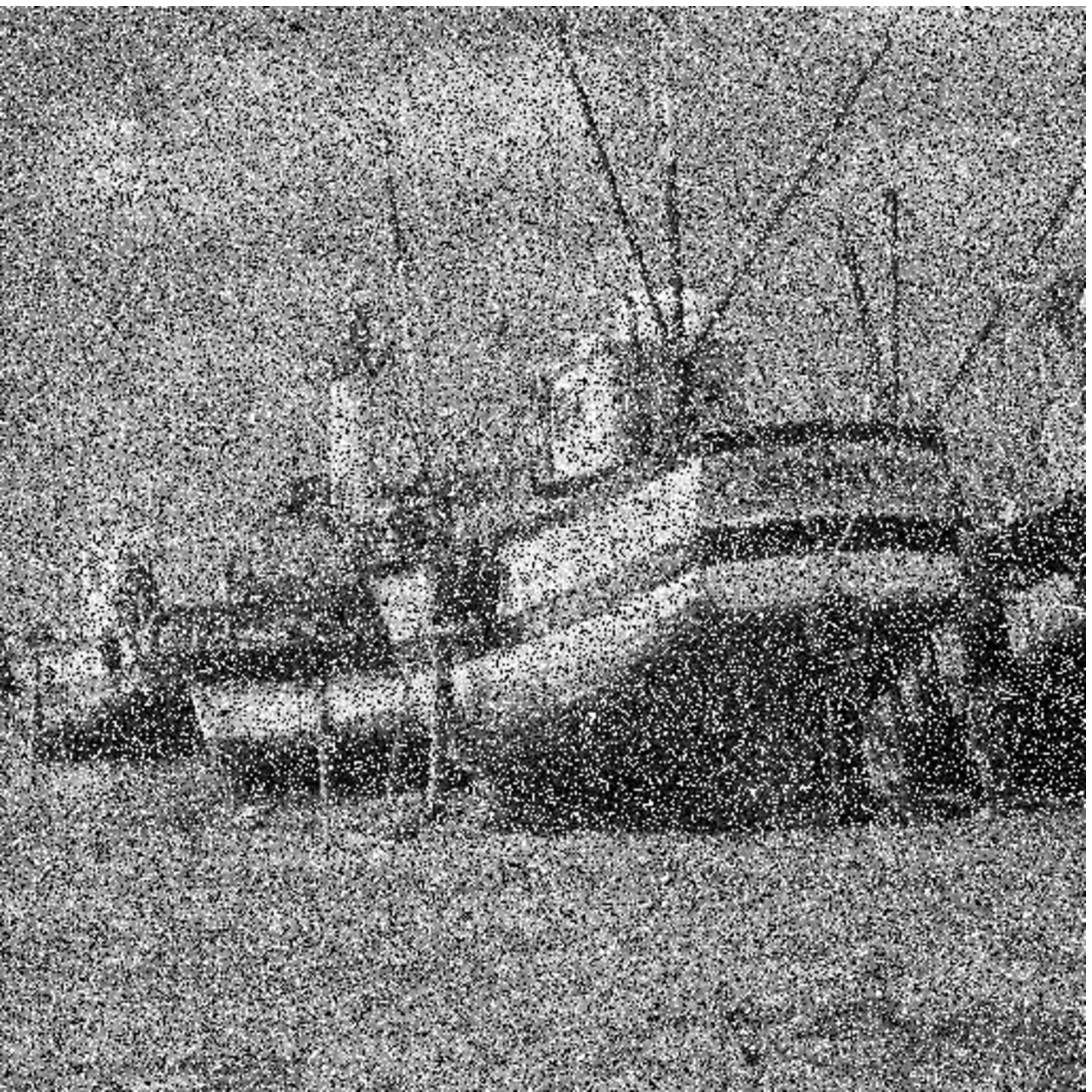}
\includegraphics[height=4cm,width=4cm]{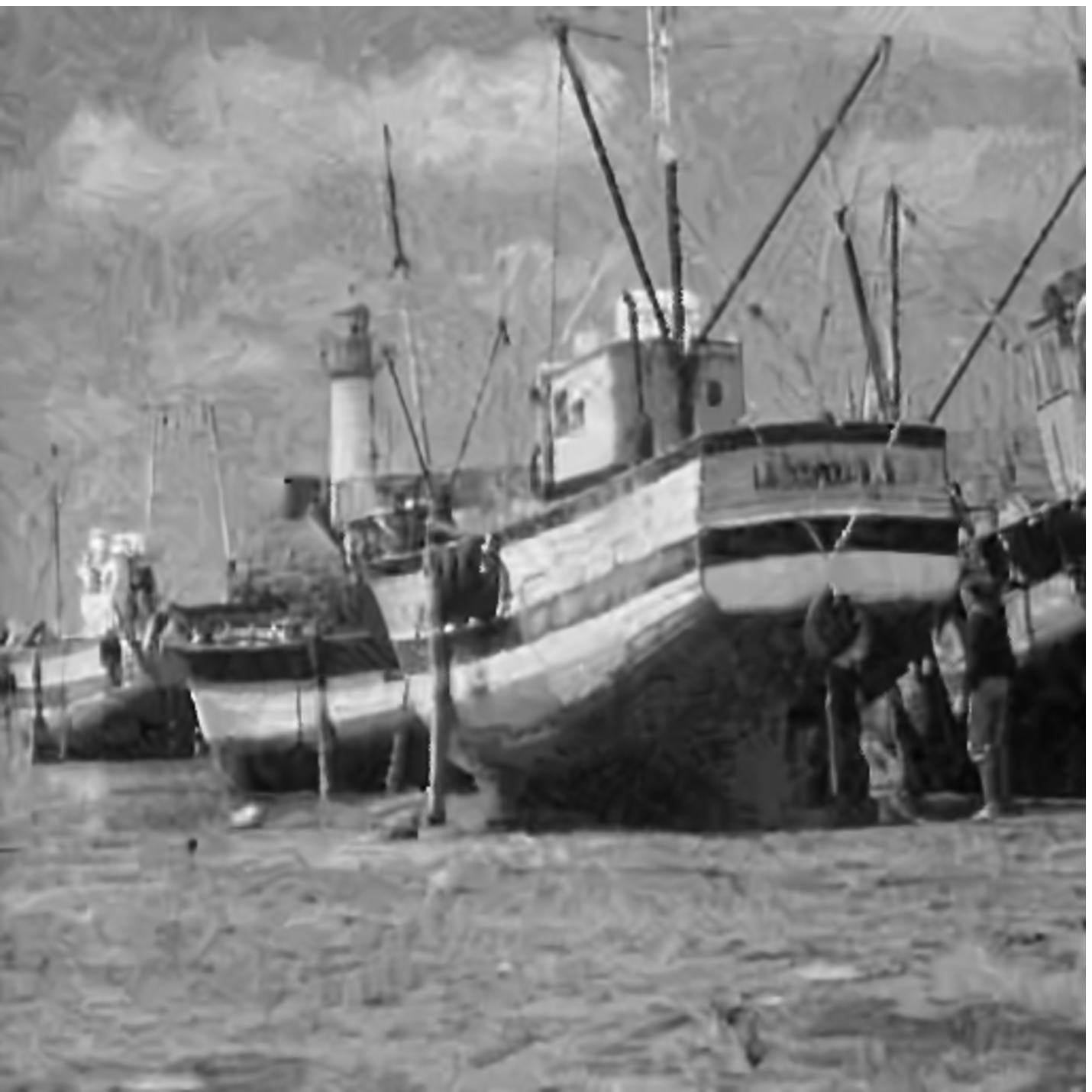}
\includegraphics[height=4cm,width=4cm]{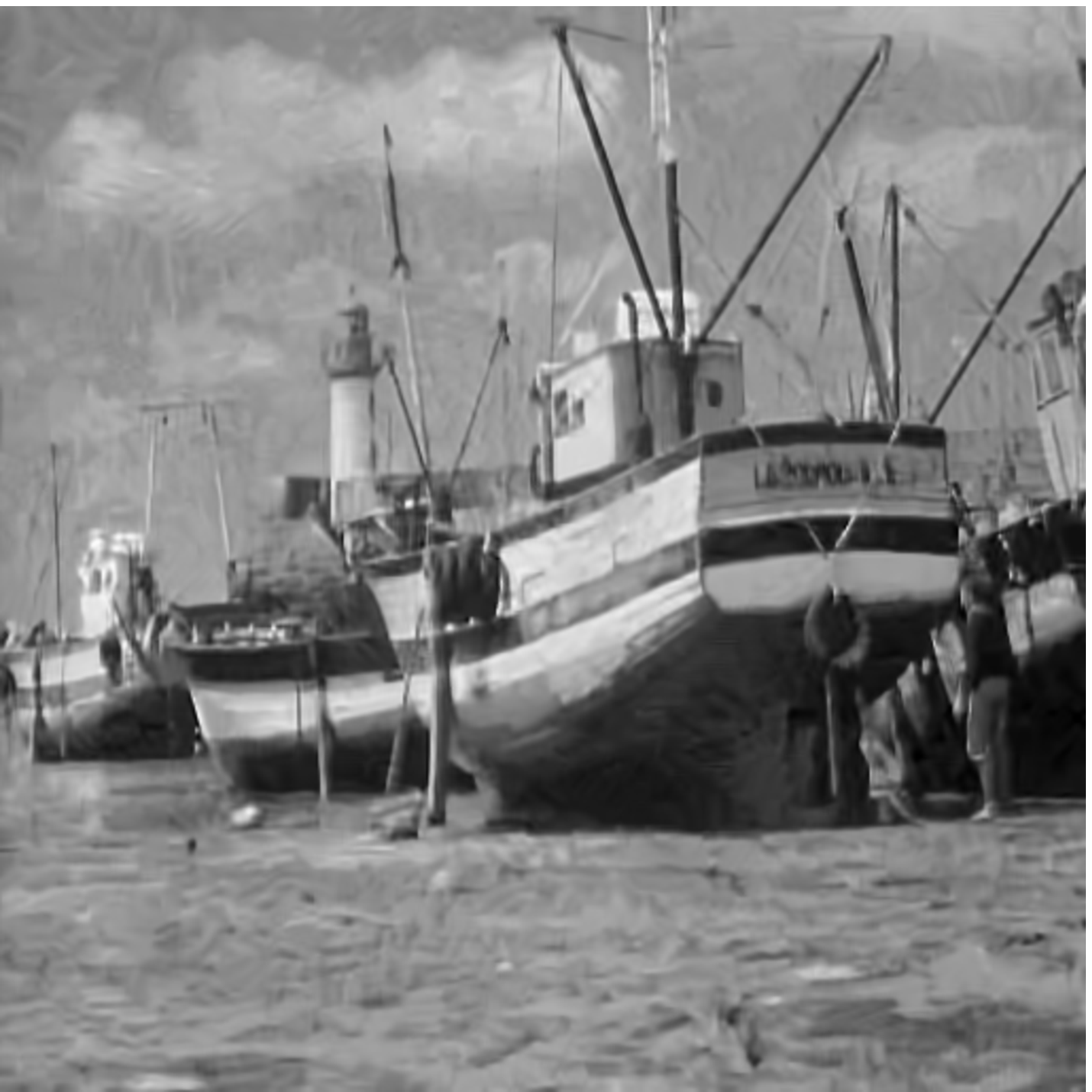}\\
(a)\hspace{4.00cm}(b)\hspace{4.00cm}(c)\hspace{4.00cm}(d)\\
\includegraphics[height=4cm,width=4cm]{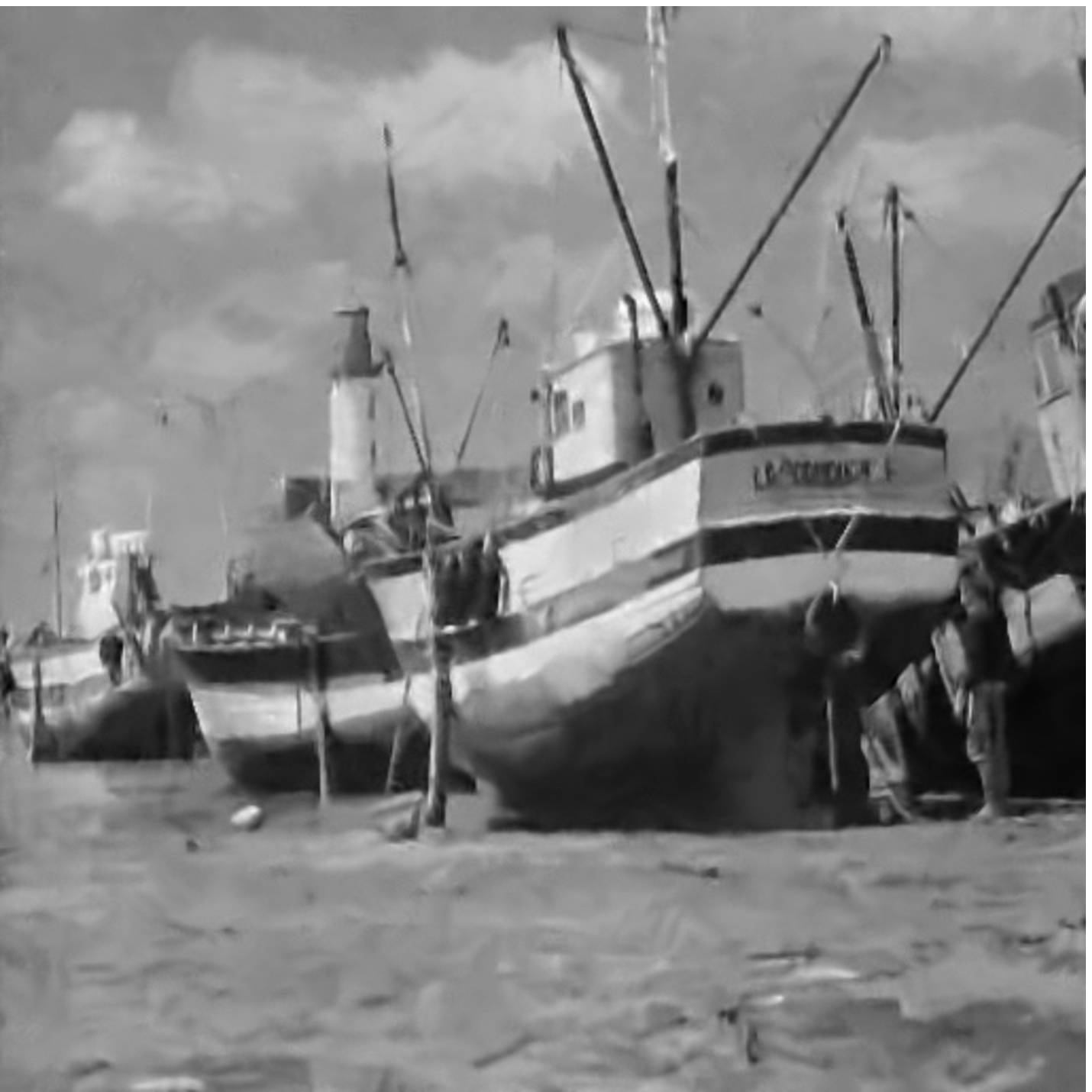}
\includegraphics[height=4cm,width=4cm]{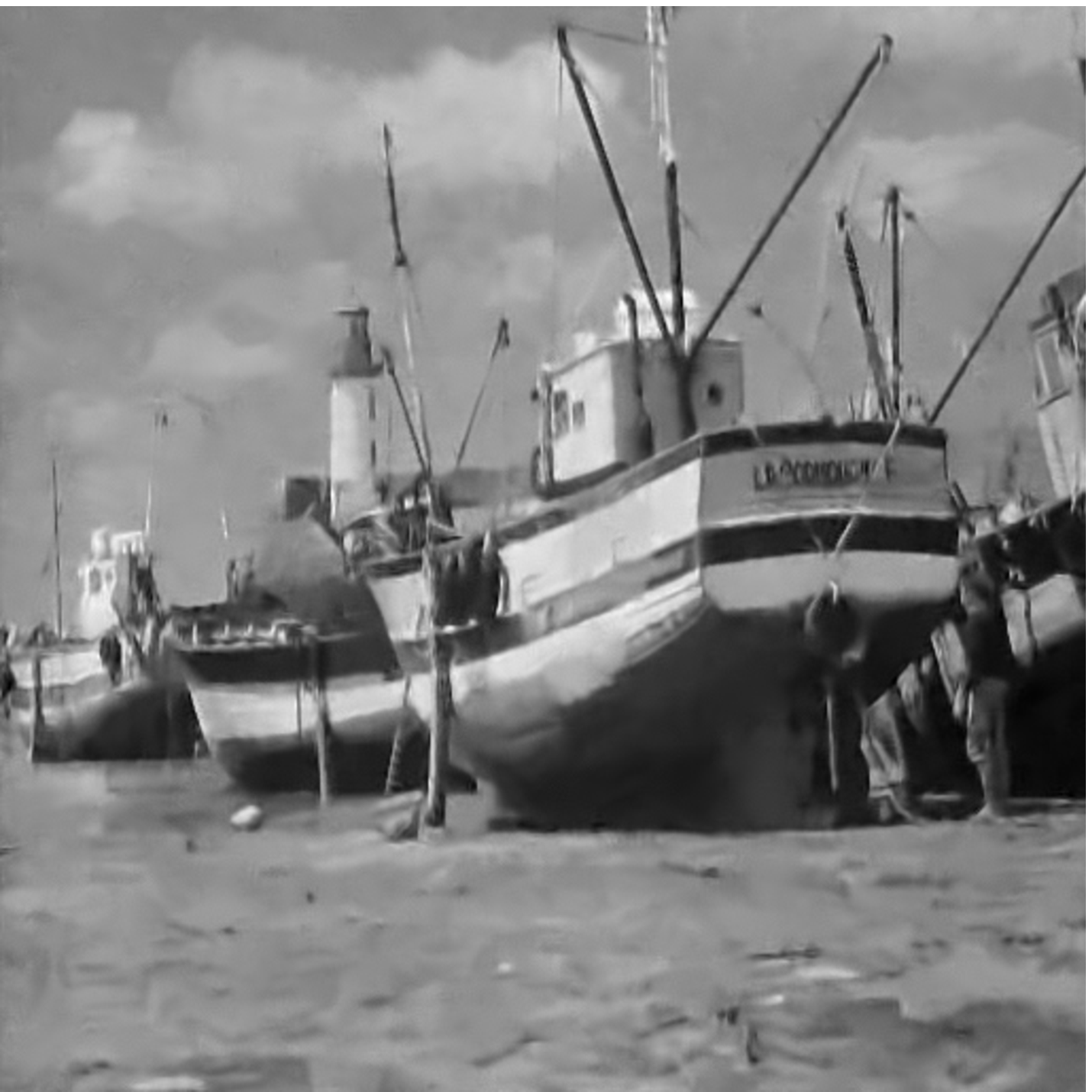}
\includegraphics[height=4cm,width=4cm]{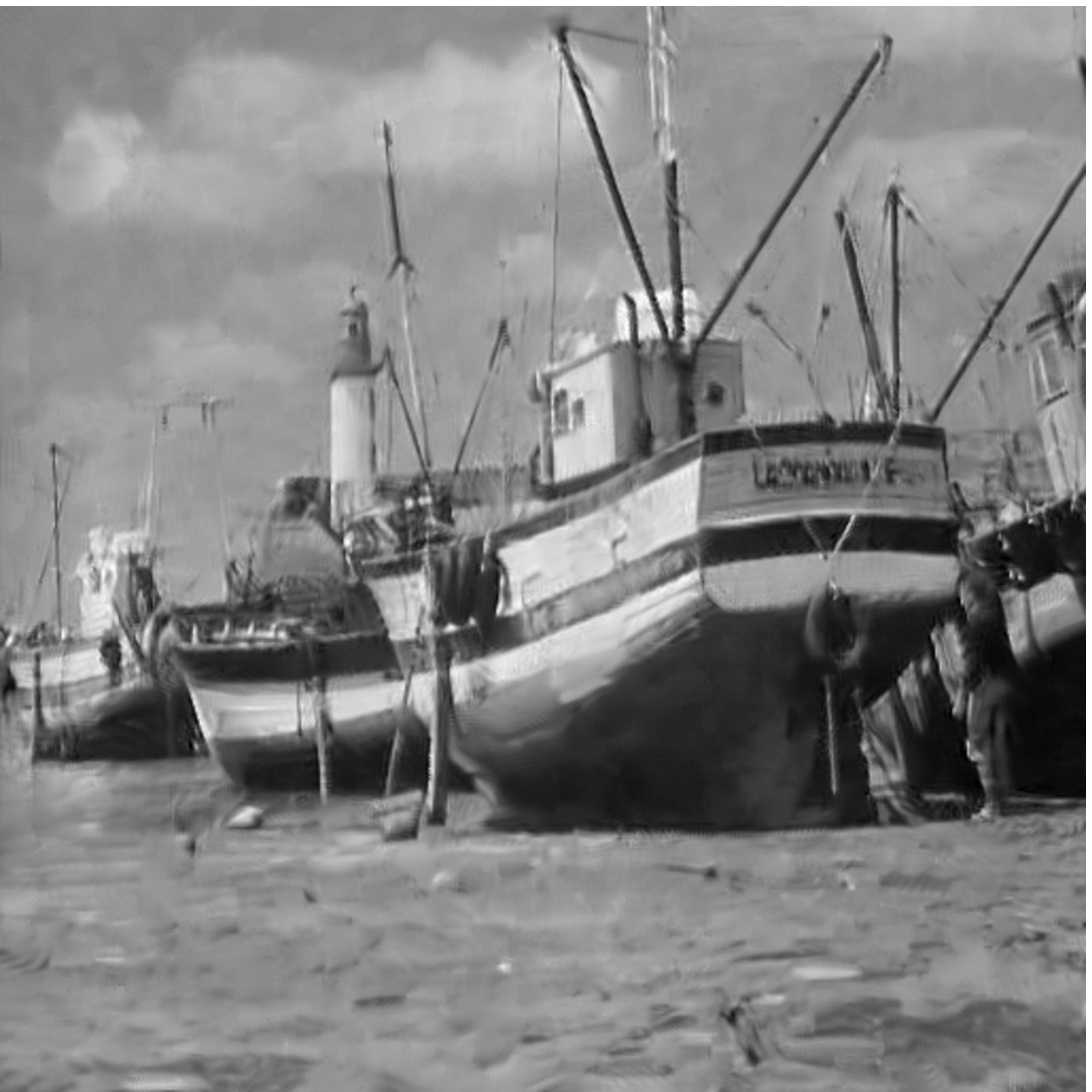}
\includegraphics[height=4cm,width=4cm]{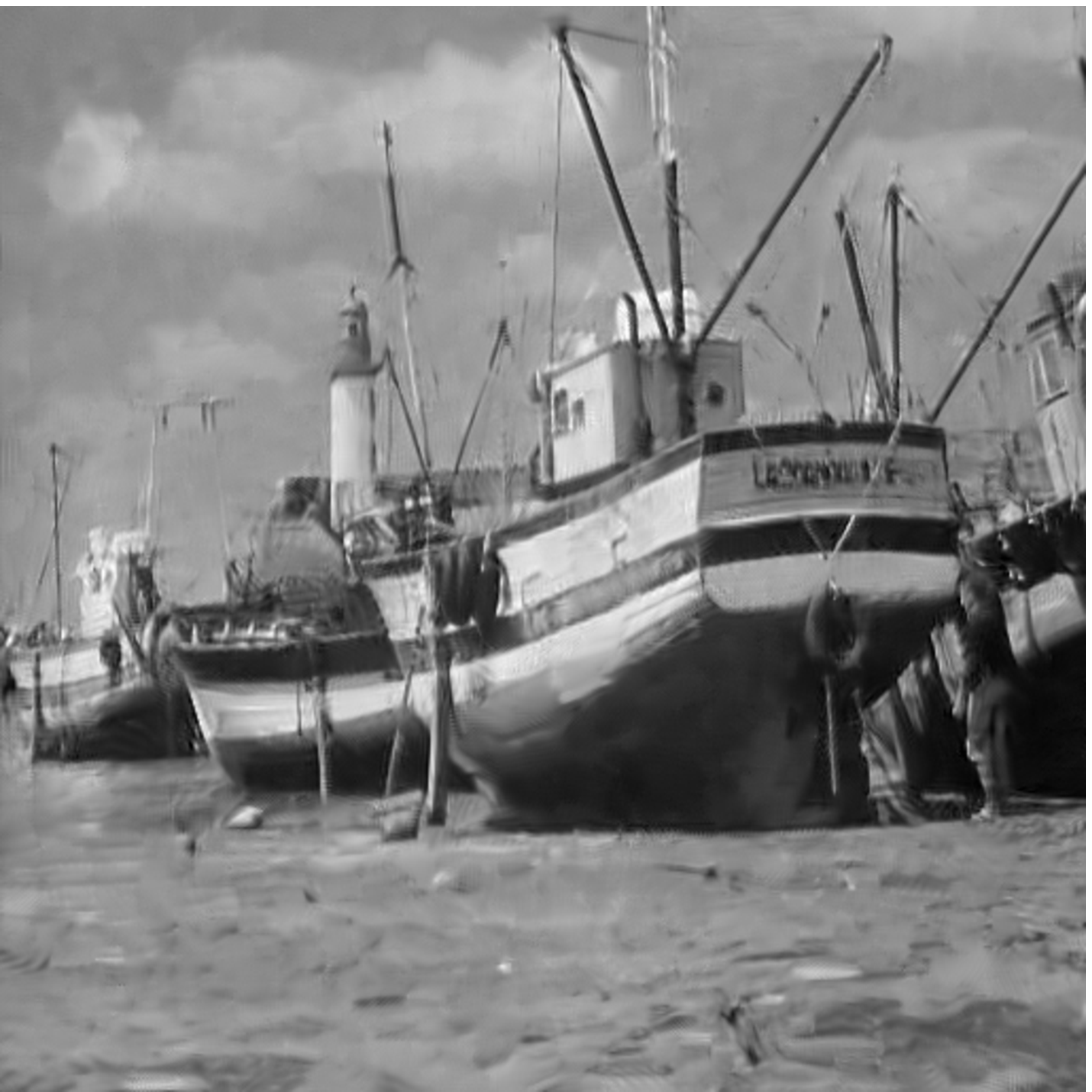}\\
(e)\hspace{4.00cm}(f)\hspace{4.00cm}(g)\hspace{4.00cm}(h)\\
\caption{\label{fig:spin_rem}Visual comparison of denoising performance of the methods for the \textit{Boat} image for mixed AWGN+SPIN removal. (a) The ground truth image. (b) The image corrupted by mixed AWGN+SPIN with parameters $\sigma=25$ and $p=30\%$ (PSNR=$10.37$ and SSIM=$0.0671$). The image recovered using (c) AMF+BM3D~\cite{dabov2007image} (PSNR=$27.22$ and SSIM=$0.7223$), (d) MBM3D (PSNR=$28.50$ and SSIM=$0.7607$), (e) WESNR~\cite{jiang2014mixed} (PSNR=$27.23$ and SSIM=$0.7230$), (f) MWESNR (PSNR=$28.06$ and SSIM=$0.7376$), (g) LSM-NLR~\cite{huang2017mixed} (PSNR=$28.56$ and SSIM=$0.7668$), and (h) MLSM-NLR (PSNR=$28.71$ and SSIM=$0.7684$).
}
\end{figure*}


    
    

In order to conduct experiments, three commonly referred methods have been chosen. The methods are: ROF+BM3D~\cite{dabov2007image}, WESNR~\cite{jiang2014mixed} and LSM-NLR~\cite{huang2017mixed}. The codes of these methods and the variational optimizer~\cite{cai2010fast} have been downloaded from respective authors' website. These methods use ROF as the mandatory first step. For experimentation regarding the proposed approach, the ROF step of these methods have been replaced by ROF followed by the variational step. The codes of the experiments have been made available in~\cite{codesvstep}. In order to differentiate between the original method and the proposed modification, an M is added as a prefix in the modified methods' name. The ROF used for mixed AWGN+SPIN removal is AMF, and for mixed AWGN+SPIN+RVIN the ROF is AMF followed by adaptive center weighted median filter (ACWMF) as in ~\cite{huang2017mixed}. The value of $\beta$ is set to $0.0002$ when only SPIN is present and $0.002$ when RVIN is present in the noise. The performance of the algorithms are measured using the peak signal to noise ratio (PSNR) and the structural similarity (SSIM) indices~\cite{wang2004image}.

Table~\ref{tab:spin} shows results of mixed AWGN+SPIN denoising on three typical images, \textit{Lena}, an image with smooth details, \textit{House}, an image with repetitive structures and \textit{Boat}, an image with higher details and with many edges in different directions. The experiment is conducted in two scenarios where the AWGN noise parameters $\sigma$ is set to $25$ and the SPIN noise parameter $p$ is varied between $30\%$ and $50\%$. Each of the methods has been run five times and the average metric is reported in the table. It can be observed from the table that the proposed modification certainly facilitates the overall denoising performance. The Gaussian denoising algorithm BM3D achieves the largest increase of performance in terms of both metrics. The WESNR and LSM-NLR method, tailored for mixed noise, also achieve significant performance boost. It can be seen that as the SPIN noise is increased, keeping the AWGN noise fixed, the percentage increase of performance also increases, which shows the proposed method is an inevitable step for high quality denoising performance for these methods. Similarly, table~\ref{tab:rvin} shows the result of mixed AWGN+SPIN+RVIN removal for $\sigma$, $p$, and $r$ set to $10$, $25\%$ and $5\%$, respectively. It can be observed that the variational step improves the denoising performance in all of the considered methods.

Fig.~\ref{fig:spin_rem} shows the visual comparison of the denoising performance of the methods for noise parameters $\sigma=25$ and $p=30\%$ using both default and proposed modification using the variational step. It can be observed from the figures that the proposed modification improves the overall quality of the images for the considered methods. The over all consistency of the images and sharpness of the edges have been improved using the proposed variational step in mixed noise removal.

\section{Conclusion}~\label{sec:conclusion}
The removal of mixed AWGN-IN is a challenge as the two types of noise are contrasting. The IN in the distribution causes a heavy tail which is hard to capture using the denoising algorithms. In order to reduce this heavy tail, in this paper, a variational method based approach has been taken. It has been shown that by introducing a variational denoising step, the heavy tail caused by the IN can be efficiently reduced. By conducting experiments, it has been observed that the proposed variational step to modify the existing methods can be employed to reduce mixed Gaussian-impulse noise with improved denoising performance. Furthermore, this approach can be adopted as an integral step in any mixed AWGN-IN removal method.


%

\bibliographystyle{IEEEtran}
\bibliography{biblio}
%



\end{document}